\documentclass[useAMS]{mn2e}

\newcommand{\msol}{M_\odot}

\newcommand{\mdot}{\dot M}
\input psfig
\title[SXT outbursts]
      {The origin of the rebrightening in soft X-ray transient outbursts}
\author[M. R. Truss et al]
       {M. R. Truss\thanks{E-mail : mtr@star.le.ac.uk}, G. A. Wynn, J. R. Murray and A. R. King\\ 
Department of Physics and Astronomy, University of Leicester, University Road, Leicester, LE1~7RH, UK}

\begin{document}

\maketitle

\begin{abstract}
We present a model of an outburst of the soft X-ray transient A0620-003. A two-dimensional time-dependent
smoothed particle hydrodynamics scheme is used to simulate the evolution of the accretion disc 
through a complete outburst. The scheme includes the full tidal potential of the binary and a simple treatment
of the thermal-viscous disc instability. In the case where the mass accretion rate onto the primary determines 
the fraction of the disc that can be kept in a hot, high viscosity state by the resulting X-ray emission, we find 
that the shape of the X-ray light curve is ultimately determined by the relative sizes of the irradiated and 
unirradiated parts of the disc and the growth time-scale of the tidal instability. The model accounts for 
the rebrightening that has been observed in the light curves of A0620-003 and several other transients. The 
primary maximum and subsequent decline are due to the accretion of gas within the irradiated portion of the disc, 
while the secondary maximum is caused by the accretion of gas in the outer part of the disc that is initially 
shadowed from the central X-rays, but subject to tidal forces. We propose that tidal effects at the disc edge can be 
sufficient to drive accretion on a time-scale shorter than that expected for a standard alpha-viscosity disc. The 
final decay is subsequently controlled by the gradual retreat of the irradiated portion of the disc. If the entire disc 
is kept in the high-viscosity state by the irradiation, no rebrightening is possible.
\end{abstract}


\section{Introduction}
The soft X-ray transients (SXTs) are close interacting binary systems consisting of a black hole or neutron 
star primary and a low-mass Roche-lobe-filling secondary star. The outbursts of soft X-ray transients are 
quite distinct from those seen in other binary systems such as dwarf novae. The duration and recurrence 
times are both strikingly longer, suggesting that a large fraction of the mass of the accretion disc is consumed 
in each outburst and that a different mechanism from the standard disc instability picture is required 
(comprehensive reviews of the disc instability picture are given by Cannizzo, 1993 and Lasota, 2001). 

Typical ionization instability-driven dwarf nova outbursts last a few days and consume a few percent of the 
disc mass, leading to a recurrence time of the order of weeks. In comparison, a typical SXT outburst lasts for
100 days or more and recurs on a time-scale of several years. The profiles of the X-ray light curves of SXT 
outbursts vary greatly but they do share common characteristics - the initial peak is often followed by an 
exponential decline and, in many cases, a rebrightening several tens of days after the first maximum. Smaller 
rebrightenings later on in the decay from outburst have also been observed. These features can be seen quite 
clearly in the X-ray light curve of the well-known transient A0620-003 (Elvis et al. 1975, Whelan et al. 1977). 
A general review of SXT outbursts is given by Tanaka \& Lewin (1996).   

Several authors have proposed models to explain the form of the SXT light curves. The long recurrence time 
of the outbursts is a problem which has been addressed in many ways. It can be explained by the standard 
ionization instability model if the Shakura-Sunyaev viscosity parameter $\rm \alpha$ is greatly reduced 
from that typical of a cataclysmic variable disc. Such an explanation is not confined to SXTs; it has been 
invoked to explain the outbursts of WZ Sagittae (Smak, 1993). However, there is no obvious physical reason 
why $\rm \alpha$ should vary so much from one accretion disc to another. The more successful models 
involve the removal of the inner part of the disc where the viscous time-scale is short. This can been achieved 
by a hot, advection-dominated accretion flow (ADAF) onto the hole : Meyer \& Meyer-Hofmeister (1994) 
showed that gas from the cool outer accretion disc in a binary system may evaporate into a hot corona and
subsequently feed an ADAF onto the comapct object. Dubus et al. (2001) have performed one-dimensional 
simulations of an SXT including the effects of both irradiation and an ADAF. In this work we do not
address the question of recurrence time, but concentrate on the observed outburst profiles. Cannizzo, Chen \&
Livio (1995) found that the exponential decay could be reproduced by parameterising $\rm \alpha$ in terms
of the aspect ratio $\rm \frac{H}{R}$. Augusteijn, Kuulkers \& Shaham (1993) sought to explain the 
rebrightenings by considering irradiation of the secondary as a mechanism for increasing the mass transfer 
rate. However, it is unlikely that the radiation can penetrate the photosphere of the companion to produce a 
variation in mass transfer rate of more than a few percent (King, 1989). 

We consider the effect of disc irradiation by the X-ray emission from the compact object. King \& Ritter 
(1998) showed analytically that the assumption that the central X-ray  source can efficiently irradiate the 
disc leads automatically to long exponential X-ray decays with the observed time-scales. Observations of 
SXTs suggest that the optical/X-ray flux ratio for these objects is simply too large for the optical emission to 
result from intrinsic viscous dissipation in the disc alone (van Paradijs \& McClintock, 1994; Shahbaz \& 
Kuulkers, 1998). Moreover, the optical brightness is strongly correlated with the irradiating X-ray flux. 
However, detailed one-dimensional modelling of the disc structure (e.g. Cannizzo 1994) suggests that the 
response of the disc to the irradiation is for the surface to become convex rather than concave, thus 
shielding the outer regions from the irradiating flux. The observations seem to contradict this picture, and for
the purposes of this paper we assume that the disc surface remains concave. 

In section 2 we describe the irradiated disc model of King and Ritter in more detail, and outline the 
importance of tidal forces in systems with extreme mass ratios in section 3. The numerical method used to 
implement the model is then described in section 4. In section 5, we show that the observed peak luminosity 
of the 1975 outburst of A0620-003 almost certainly implies that the irradiation radius (the outer radius of the 
region kept heated by the central X-rays) was larger than the accretion disc. We then show in simulations 
that no rebrightening occurs if the whole disc is kept in the hot viscous state by the irradiation; however, if only
a very small area of the outer disc is shielded in some way (by a radiation-induced warp or self-shadowing,
for example), the observed rebrightening is produced.
\section{Disc Irradiation Model}
Irradiation of an accretion disc by the central X-ray source heats the exposed disc faces. In SXTs, provided 
the disc faces remain concave, the surface temperature of the disc is given by the formula of de Jong, van 
Paradijs \& Augusteijn (1996) : 
\begin{equation}
\rm{ T_{irr}^{~4}(R) = {{\eta \dot{M}c^{2}(1-\beta)} \over {4\pi\sigma R^{2}}} \left({H \over R}\right)^{n} \left[{d~ln H \over d~ln R} - 1\right]}
\label{tirr}
\end{equation} 
\noindent
where $\rm \eta$ is the accretion efficiency for producing X-rays, \.M  is the mass accretion rate onto the 
primary, H is the scale-height of the disc and $\rm \beta$ is the disc albedo. The index n takes the values
1 or 2 for irradiation by a point source or by the inner disc respectively. Accretion onto a neutron star will 
always have n = 1, while n = 2 is possible in the absence of a hard surface on which to accrete, i.e. a black hole 
primary. During outburst, however, the disc will probably develop an X-ray corona, as suggested by 
observations of a strong power-law continuum. In this case, the appropriate index reverts to n = 1. 
In a large enough disc, $\rm T_{irr}$ is quite sufficient to dominate the effective temperature and exceed the
hydrogen ionization temperature ($\rm T_{H} \sim 6500 K$), thus stabilising the disc on the hot branch of 
the $\rm T_{eff} - \Sigma$ relation (see for example, Cannizzo, 1993). $\rm T_{irr}$ is the key factor
which determines whether a low mass X-ray binary is transient or not - in a persistent source $\rm T_{irr}$
exceeds $\rm T_H$ throughout the entire disc (van Paradijs, 1996).
\begin{figure}
\psfig{file=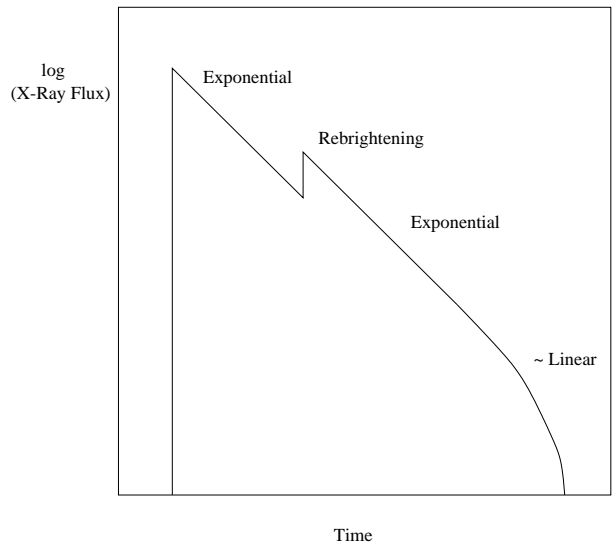,width=8cm}
\caption{Schematic representation of a SXT outburst according to the King \& Ritter (1998) model.}
\label{scheme}
\end{figure}

We consider a situation in which the disc undergoes an outburst which is initiated by the standard hydrogen 
ionization disc instability. Subsequently, as the mass accretion rate onto the primary increases, the disc is 
irradiated by the enhanced X-ray emission. The disc is unable to return to a cool quiescent state until the mass
accretion rate has fallen sufficiently to switch off the enhanced irradiation, allowing a cooling front to sweep 
through the disc. Following the analytic treatment of King \& Ritter (1998), one can use equation 1 to 
determine the variation of $\rm R_h$ with $\rm \dot M_c$ by noting that the edge of the irradiated region 
is determined by the point in the disc at which $\rm T_{irr} = T_H$. Then :
\begin{equation}
\rm{ R_h^{2} = {{\eta \dot{M}c^{2}(1-\beta)} \over {4\pi\sigma {T^4}_H}} \left({H \over R}\right)^{n} \left[{d~ln H \over d~ln R} - 1\right] = B_n \dot M}
\label{rh}
\end{equation} 
\noindent
Using typical parameters for neutron star and black hole systems, King \& Ritter showed that the mass 
accretion rates for a given $\rm R_h$ are then
\begin{equation}
\rm{ \dot M_{1,crit} = 4.1 \times 10^{-10} {R_{11}}^2 \msol yr^{-1},}
\label{ns}
\end{equation}
\noindent
and
\begin{equation}
\rm{ \dot M_{2,crit} = 2.9 \times 10^{-9} {R_{11}}^2 \msol yr^{-1},}
\label{bh}
\end{equation}
\noindent
where $\rm R_{11}$ is $\rm R_h$ in units of $\rm 10^{11}~cm$. King and Ritter and subsequently 
King (1998) extend this analysis to show that the first part of the outburst decay, with the whole disc 
irradiated, will be exponential, but ultimately the final part of the decay will be roughly linear due to the 
gradual shrinking of the irradiated region. However, they could not fully explain the rebrightening, and its 
origin remained unknown. A schematic of the theoretical profile is given in fig.~\ref{scheme}.

The recurrent transient A0620-003 (V616 Mon) is perhaps the best known of the SXTs. Its 1975 outburst 
was observed by Ariel V (Elvis et al. 1975), during which second and third maxima were observed 55 and 200
days after the first. The decay either side of the first rebrightening is exponential and the rise-time of the first
rebrightening  was $~\sim 10 - 15$ days. A0620-003 has a short orbital period of 7.75 hours (McClintock et al., 
1983), and is likely to contain a black hole primary. The primary mass has been estimated to be 
$\rm 7.3 \msol$ or more (McClintock \& Remillard, 1986), but a more recent work concludes that it is likely
to be in the range $\rm 3.30 < M_1 <4.24~\msol$  (Marsh,  Robinson \& Wood 1994). The reader is referred
to Whelan et al. (1977) or Tanaka \& Lewin (1995) for the full X-ray light curve of the 1975 outburst. 

The analysis of King \& Ritter correlates well with the observed behaviour of A0620-003 and SXTs in general,
but the precise origin of the rebrightening and the behaviour of the disc remain to be explained. In this work 
we incorporate King \& Ritter's analytic model into an existing numerical code and perform two-dimensional
simulations of a SXT with the orbital parameters of A0620-003. We find a natural explanation for the 
rebrightening in terms of the physical state of the disc. In a forthcoming paper we will report results from
simulations of outbursts in longer period SXTs (with correspondingly larger discs) and investigate the 
resulting complex outburst morphologies.

\section{Tides}
The mass ratio of A0620-003 is very well-constrained :  $\rm q = {M_2 \over M_1} = 0.067 \pm 0.01$  
(Marsh. Robinson \& Wood, 1994). Such a low value suggests that tidal resonances will be accessible to the 
disc. Tidal forces will therefore be an important factor in the dynamics of the accretion disc. The radius of the 
3:1 tidal resonance, which has been shown to be important in dwarf novae,  is 
\begin{equation}
\rm{{R_{3:1} \over a} = \left ( {1 \over 3 }\right )^{2 \over 3}(1 + q)^{-{1 \over 3}}}
\label{rtide}
\end{equation}
\noindent
where a is the binary separation (see, for example, Frank, King \& Raine, 2002). For $\rm q = 0.07$,  this 
radius is $\rm R_{3:1} = 0.47a$. This is well inside the radius at which the disc is ultimately truncated by the 
tides. $\rm R_{tide}$ can be estimated from the Roche lobe radius of the primary - it is always roughly 90\% 
of this value. Using the formula of Eggleton (1983),
\begin{equation}
\rm{ {R_L(1) \over a} = {0.49 q^{-{2 \over 3}} \over {0.6 q^{-{2 \over 3}} + log_e (1 + q^{-{1 \over 3}})}}}~~~~,
\label{rl1}
\end{equation}
\noindent
$\rm R_{tide} \simeq 0.60a$ for \rm q = 0.07. Therefore we expect tidal effects to become important and  the 
disc to become eccentric when it reaches  $\rm R_{3:1}$.
\section{Numerical Method}
We use a two dimensional smoothed particle hydrodynamics (SPH) accretion disc code. The great advantage
of this method is that it incorporates the full tidal potential of the binary system without need for an 
approximate prescription for the removal of angular momentum at the disc edge employed in 1-D 
simulations. The importance of using a two- or three-dimensional scheme with the tidal effects included has 
been demonstrated effectively with this code in recent simulations of cataclysmic variables (Truss et al. 2001).
Full details of the code are described in Truss et al. (2000) and a complete review of the SPH technique is given
by Monaghan (1992). The implementation here is very closely related to the dwarf nova study of Truss et al. 
(2000,2001), in which we incorporate the disc instability in terms of a viscosity switch based on surface density
and build up the disc by a mass transfer stream. The disadvantage of the scheme is that there is no 
self-consistent calculation of vertically-averaged thermal structure, as in one-dimensional codes. There is a 
trade-off in computational efficiency between calculating the thermodynamics in a self-consistent manner in 
1-D and performing the hydrodynamics in 2-D. We choose to present a 2-D calculation in this work to 
examine the role of tidal forces, and in this way the scheme is complementary to modern 1-D treatments.
\subsection{Viscosity}
When the local surface density at the position of any particle in the disc exceeds the critical value for 
ionization of hydrogen to occur as calculated by Cannizzo,  Shafter \& Wheeler (1988), 
\begin{equation}
\rm{ \Sigma_{max} =
11.4R_{10}^{1.05}M_{1}^{-0.35}\alpha_{cold}^{-0.86} ~gcm^{-2} },
\label{trig1}
\end{equation}
\noindent
the viscosity acting on that particle is increased. The change  in viscosity is made on an appropriate thermal 
time-scale (much shorter than the local viscous time-scale)  from a value $\rm \alpha_{cold}$ to a value 
$\rm \alpha_{hot}$, where $\rm \alpha$  is defined in terms of the sound speed of the fluid, $\rm c_s$, and 
the scale height of the disc, H, in the usual viscosity formulation of Shakura \& Sunyaev (1973):
\begin{equation}
\rm{ \nu = {2 \over 3}\alpha c_s H}.
\label{ss}
\end{equation}
\noindent
The local viscosity remains at $\rm \alpha_{hot}$ until the local surface density decreases below a critical 
value 
\begin{equation}
\rm{ \Sigma_{min} =
8.25R_{10}^{1.05}M_{1}^{-0.35}\alpha_{hot}^{-0.8} ~gcm^{-2}}. 
\label{trig2}
\end{equation}
\noindent
when it is returned to the low state value $\rm \alpha_{cold}$.  

Equation~\ref{ss} is employed almost directly in the SPH formulism. The SPH momentum equation for a 
particle i is:
\begin{equation}
\rm {d{\bf v}_i \over dt}=-\sum_jm_j\left({P_i \over \rho_i^2}+{P_j \over \rho_j^2}+{\beta\mu_{ij}^2-\zeta\bar c_{ij}\mu_{ij} \over \bar \rho_{ij}}\right)\nabla_iW_{ij}
\label{mom}
\end{equation}
\noindent
where $\rm \bar \rho_{ij}$ and  $\rm \bar c_{ij}$ are the average density and sound speed of the pair of 
particles i and j respectively. h is the smoothing length and $\rm W_{ij}$ is the interpolating kernel. 
$\rm \beta$ and $\rm \zeta$ are the quadratic and linear artificial viscosity parameters and
\begin{equation}
\rm  \mu_{ij} = \left \{ \begin{array}{ll} 
\frac{{\bf v_{ij}} \cdot {\bf r_{ij}}}{{\bf r_{ij}}^2 + \eta^2} & \mbox{$\rm {\bf v}_{ij} \cdot {\bf r}_{ij} \leq 0$} \\ 
0 & \mbox{$\rm {\bf v}_{ij} \cdot {\bf r}_{ij} > 0$} 
\end{array}         \right .
\label{piv}
\end{equation}
\noindent
where $\rm \eta^2 = 0.01h^2$ avoids singularities. 

Following Murray (1996),  we set $\rm \beta =0$ and in the continuum limit (i.e. $\rm h \rightarrow 0$ and
 number of particles $\rm N \rightarrow \infty$) the acceleration due to the artificial viscosity term can be written
\begin{equation}
\rm {\bf a}_v = \frac{\zeta H \kappa}{2\Sigma}(\nabla \cdot (c_s\Sigma S) + \nabla(c_s\Sigma \nabla \cdot {\bf v})),
\label{accel}
\end{equation}
\noindent
where $\rm \kappa$ is a constant determined by the kernel W. In this case we use a 2D cubic spline kernel, 
for which $\rm \kappa = \frac{1}{4}$. $\rm S_{ij}$ is the deformation tensor, where
\begin{equation}
\rm S_{ij} = \frac{\partial v_i}{\partial x_j} + \frac{\partial v_j}{\partial x_i}.
\label{tensor}
\end{equation}
\noindent
If the density and sound speed vary on lengthscales much longer than the velocity, and if we assume that the 
fluid is divergence-free,  then
\begin{equation}
\rm {\bf a}_v = \frac{\zeta H c_s}{8} \nabla^2 {\bf v}.
\label{sphv}
\end{equation}
\noindent
So, in the continuum limit, the artificial viscosity introduces a shear viscosity 
\begin{equation}
\rm \nu = {1\over8} \zeta c_s H
\label{sphvisc}
\end{equation}
\noindent
In fact, Murray (1996) used the SPH smoothing length, h, as the lengthscale in his prescription for viscosity.
We do not impose such a constraint here : the viscous energy is dissipated over the disc scale height, H.
\begin{figure}
\psfig{file=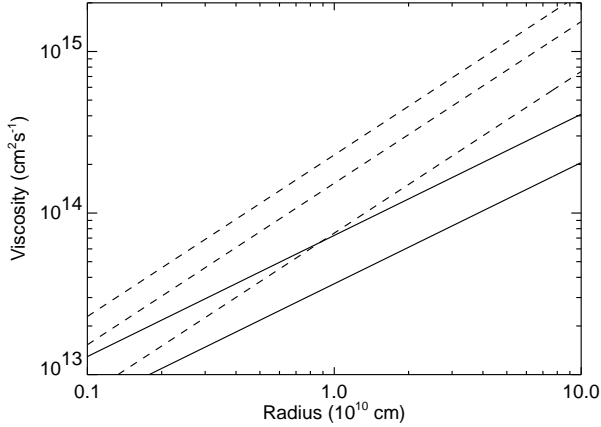,width=8cm}
\caption{Radial variation of SPH viscosity and $\rm \alpha$-viscosity. The two solid lines show the viscosity 
in an $\rm \alpha$-disc with $\rm \alpha = 0.1$ and $\rm M_1 = 4 \msol$. The upper is for 
$\rm \dot M = 10^{-8} \msol yr^{-1}$ and the lower for  $\rm \dot M = 10^{-9} \msol yr^{-1}$. The dashed 
lines show the SPH viscosity where  $\rm \zeta =1, M_1 = 4 \msol$, $\rm c_s = 0.05 a\Omega_b$ and 
$\rm \zeta = 1.5 $ (top), $\rm \zeta = 1.0$ (middle) and  $\rm \zeta = 0.5$ (bottom). a is the binary 
separation and $\rm \Omega_b$ is the angular velocity of the binary.} 
\label{sphviscplot}
\end{figure}
In the simulations, we take the sound speed to be constant everywhere and vary $\rm \zeta$ to change the 
viscosity of a particle in the way described above. With such an isothermal scheme, the net viscosity in the 
simulations has a slightly different radial dependence to an analytic alpha-disc. In the alpha model, the
variation of sound speed and scale height with radius ($\rm c_s \sim R^{-\frac{3}{8}}$ and $\rm H \sim
R^{\frac{9}{8}}$) lead to 
\begin{equation}
\rm \nu_{\alpha} = 1.8 \times 10^{14} \alpha^{\frac{4}{5}} \dot M_{16}^{\frac{3}{10}} M_1^{-\frac{1}{4}} R_{10}^{\frac{3}{4}}~~~cm^2 s^{-1}.
\label{sphal}
\end{equation}
\noindent
where $\rm \dot M_{16}$ is the mass accretion rate in units of $\rm 10^{16} gs^{-1}$.
In the SPH simualtions, the sound speed is a constant and the scale height is a linear function of radius:
\begin{equation}
\rm  \nu_{sph} = 6.4 \times 10^{14} \zeta \left ( \frac{c_s}{5 \times 10^{6} cm s^{-1}}\right )^2 R_{10}~~~cm^2 s^{-1}.
\label{sphnu}
\end{equation}
\noindent
where we usually use $\rm \zeta_{cold} = 0.1$ and $\rm \zeta_{hot} = 1.0$.
A comparison of these two viscosity variations is shown in fig.~\ref{sphviscplot}. A full discussion of how the results 
of this scheme can be interpreted in relation to observations, and its limitations, is given in Section 6.
\subsection{Radiation}
The irradiation radius is calculated by direct implementation of equation~\ref{ns} at each timestep . All 
particles within this radius are instantaneously given the high-state SPH viscosity $\rm \zeta_{hot}$. Since 
we know the number of particles being accreted onto the primary in the simulation, we can readily calculate 
the mass accretion rate at each timestep, and hence the irradiation radius. We are not concerned with 
calculating accurate optical light curves in these simulations; instead we concentrate on the total X-ray 
emission, which we take to be proportional to the mass accretion rate onto the primary. This scheme is ideal 
for investigating how variations in the X-rays arise from waves of mass travelling through the disc.\\

Our model is distinct from those employed by Huang \& Wheeler (1989) and Mineshige \& Wheeler (1989) 
in that we follow the evolution of the whole disc in two dimensions and there is no need to approximate the 
binary potential. In particular, for numerical reasons, Mineshige \& Wheeler used a very small disc in their 
simulation of A0620-003, using an outer radius of 3.16 $\rm{ \times 10^{10} cm}$, well inside even the 
circularisation radius. We simulate the entire disc and show that the outermost regions can play a pivotal 
r\^ole in determining the form of the observed X-ray light curve. 

\section{Results}
\subsection{Irradiation efficiency}
\begin{figure}
~\psfig{file=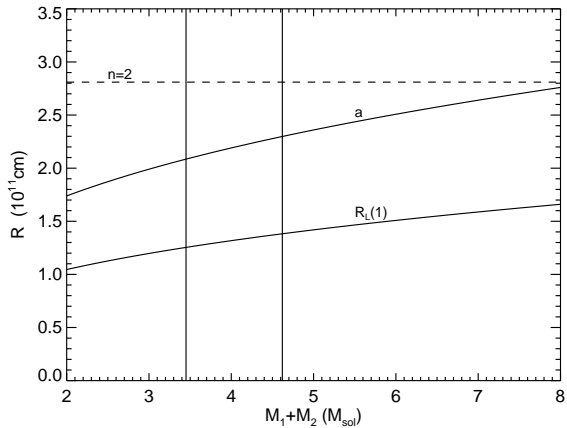,width=8cm}
\caption{The upper curve is the binary separation, a, for $\rm M_1 = 3.7 \msol$ and the lower curve is the 
Roche lobe radius of the 
primary, $\rm R_L(1)$, in A0620-003 with a mass ratio $\rm q = 0.07$. The disc radius will not exceed this 
value. The range of masses derived by Marsh, Robinson \& Wood (1994) is denoted by vertical lines and the 
dashed horizontal line shows the maximum value of $\rm R_h$ deduced from the observed peak luminosity 
of the 1975 outburst and the weaker (n=2) prescription. It is clear that $\rm R_h$ is well outside the disc and
that this constraint does not change even for masses in excess of 7$\rm \msol$. The n=1 prescription leads to
a radius of  $\rm R_h = 6.1 x 10^{11}cm$, well above the upper limit of the plot.}
\label{a0620}
\end{figure}
\noindent
Given a distance of 870 pc (Oke, 1977), the peak X-ray luminosity of the 1975 outburst of A0620-003 was 
$\rm L_x \sim 1.3 x 10^{38} erg.s^{-1}$. The total accretion luminosity can be written
\begin{equation}
\rm{ L_x = \eta \dot M c^2},
\label{masseff}
\end{equation}
\noindent
where $\rm \eta$ is the rest mass to energy accretion efficiency. We can use this equation with King \& 
Ritter's analytic expressions for $\rm R_h$ to estimate the maximum value of $\rm R_h$ reached at the
peak of the 1975 outburst. Equations (3) and (4) were derived using specific values of $\rm \eta$; (3) 
assumes that the efficiency for accretion onto a neutron star is $\rm \eta =0.15$, while for the black hole case
(4) assumes $\rm \eta =0.10$. If we relax this constraint and rewrite the equations in terms of a general 
$\rm \eta$ consistent with (\ref{masseff}), we have :
\begin{equation}
\rm \dot M_{crit} = {6.2 x 10^{-11} \over \eta}R^2_{11} ~\msol yr^{-1} \hspace{2cm}
(n=1),
\label{n1}
\end{equation}
\noindent
and
\begin{equation}
\rm \dot M_{crit} = {2.9 x 10^{-10} \over \eta}R^2_{11} ~\msol yr^{-1} \hspace{2cm}
(n=2).
\label{n2}
\end{equation}
\noindent

Now, if we infer the mass accretion rate from the observed luminosity in (\ref{masseff}), we see from 
(\ref{n1}) and (\ref{n2}) that the estimates of $\rm R_h$ are {\em independent} of $\rm \eta$. For 
n=1 we infer that $\rm R_h = 6.1 \times 10^{11}cm$ and for n=2, $\rm R_h = 2.8 \times 10^{11}cm$. For a 
total system mass of $\rm M_1 + M_2 = 4\msol$ and orbital period 7.75 hours, the binary separation of the 
system is $\rm 2.2 \times 10^{11}cm$. Therefore, the maximum value reached by $\rm R_h$ during an 
outburst of A0620-003 is outside the disc, and quite possibly much larger than the disc. The parameter space for
varying masses is shown in figure~\ref{a0620}. Given a mass ratio, q, the disc size can be estimated  using the 
formula for the volume radius of the Roche lobe of the primary, equation~\ref{rl1}. This is shown in the lower
curve of fig.~\ref{a0620}, for q = 0.07.

\subsection{Outbursts}
We perform separate simulations of A0620-003, allowing the irradiation radius to take different values
for the same mass accretion rate. In this way we are effectively varying the parameter $\rm B_n$ in 
equation \ref{rh}. We aim to discover the effects of allowing $\rm R_h$ to become larger
than the accretion disc or not. In these simulations we take $\rm P_{orb} = 7.75 h$, $\rm M_1 = 3.7 \msol$ 
and $\rm M_2 = 0.25 \msol$, consistent with the measurements of Marsh, Robinson \& Wood (1994) and
$\rm -\dot{M}_2 = 3 x 10^{-11} \msol yr^{-1}$ after McClintock et al. (1983) and Cannizzo (1998). The 
sound speed is $\rm c_s = 0.05 a \Omega_b = 24.5 km s^{-1}$ and the high and low-state SPH viscosities are 
$\rm \zeta_{hot} = 1$ and $\rm \zeta_{cold}$ = 0.1.
\begin{figure}
\psfig{file=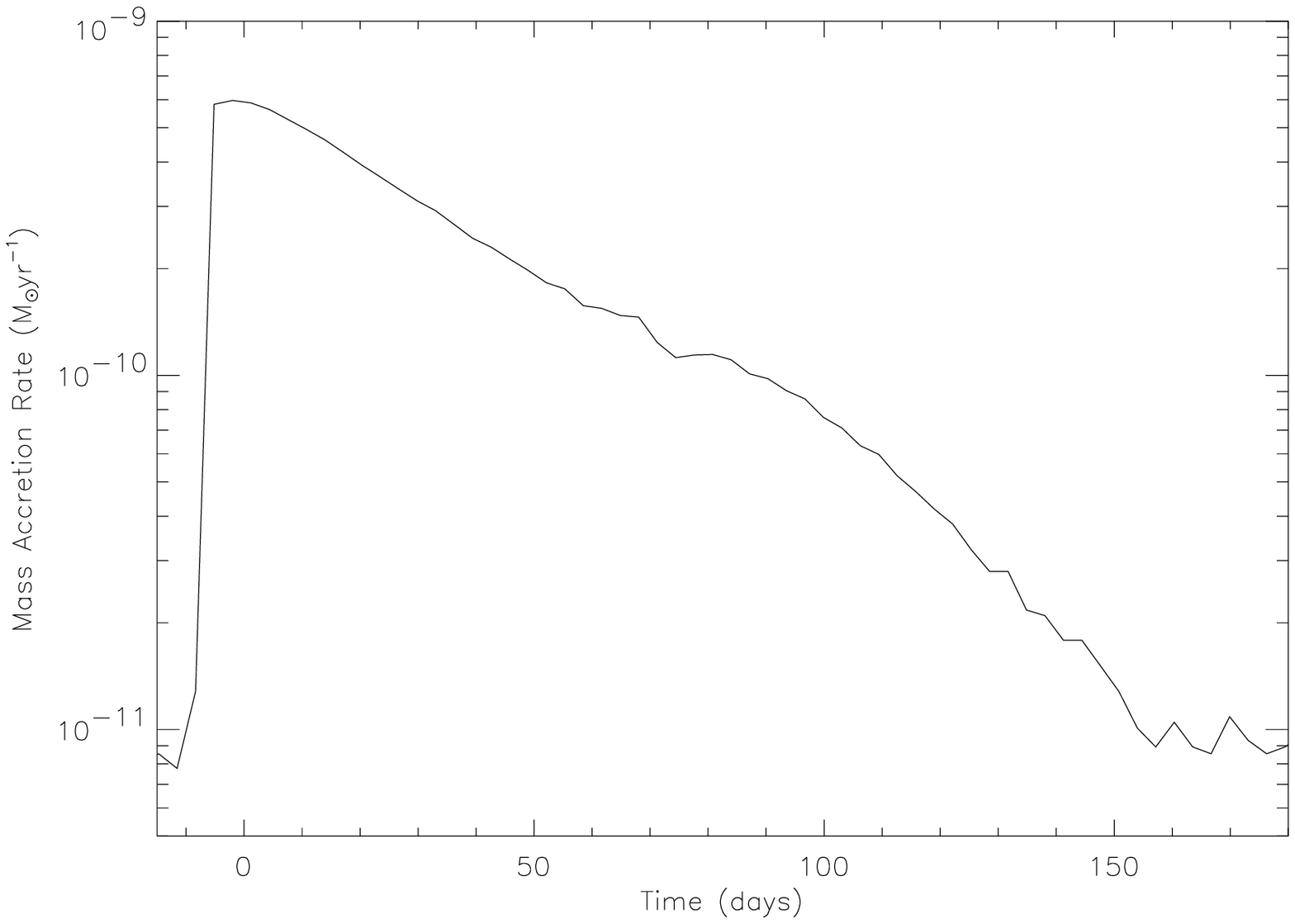,width=8cm}
\psfig{file=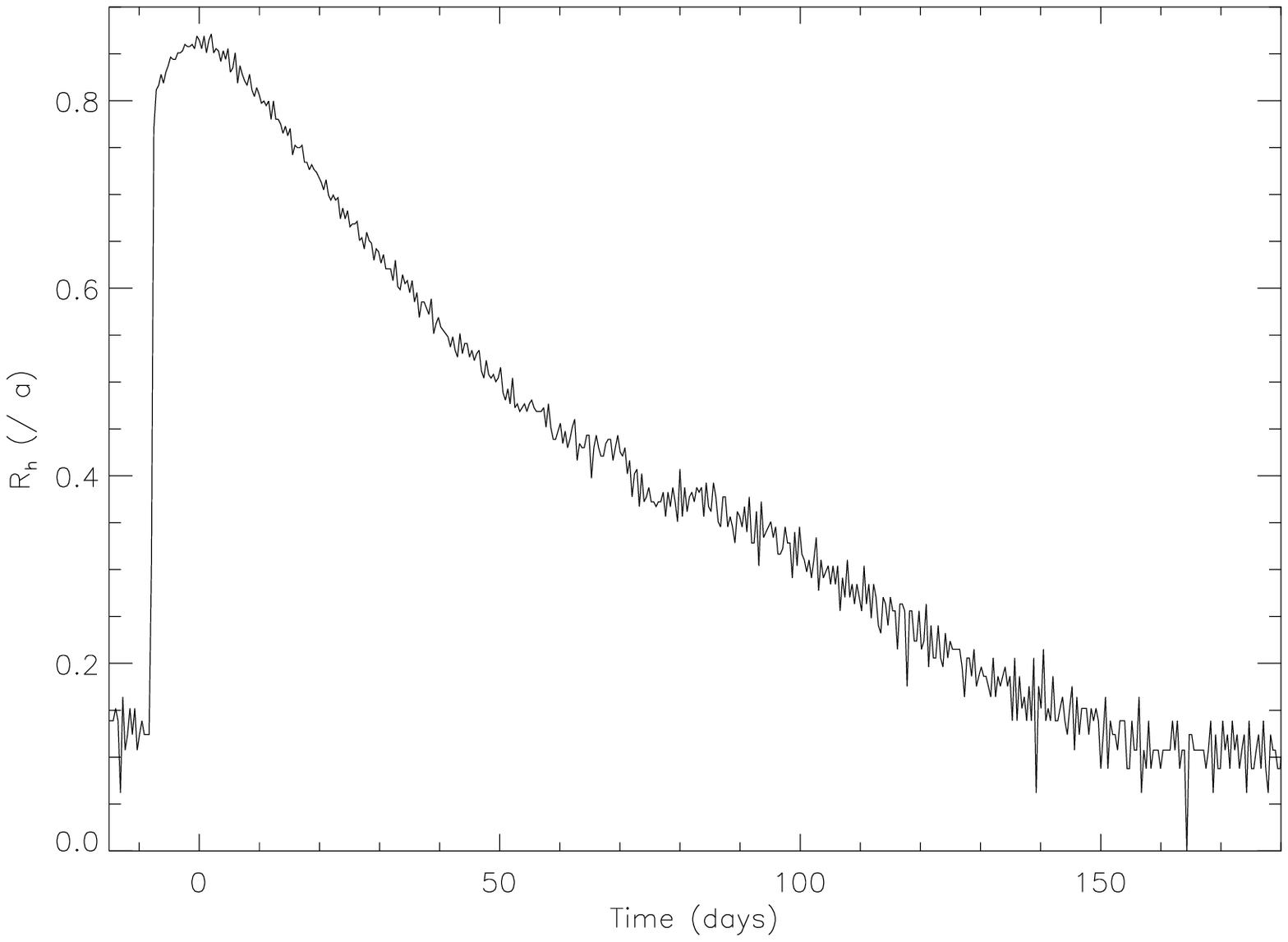,width=8cm}
\caption{Outburst simulation of A0620-003 with $\rm R_{h,max} >> R_{disc}$. The upper panel shows the 
mass accretion rate onto the primary. Below is the variation in $\rm R_h$ through the simulation. The 
decay is initially exponential, but when the irradiation retreats inside the disc, the decay becomes faster.}
\label{aobursts}
\end{figure}
\noindent

Figure~\ref{aobursts} shows the variations in mass accretion rate onto the primary and heating radius
$\rm R_h$ for the simulation of A0620-003. The entire disc is irradiated and there is no large rebrightening. 
There is an exponential decay as the irradiated gas is accreted followed by a faster decline as the accretion 
rate is controlled by $\rm R_h$ gradually retreating inside the disc. There is a small variation at 70 days 
after outburst maximum. This arises from the hot / cold boundary passing through structure near the disc 
edge associated with the spiral arms. The effect is too small to be responsible for the larger rebrightening in 
A0620-003, but may be responsible for the later, smaller glitches observed in the light curve.
\begin{figure}
\psfig{file=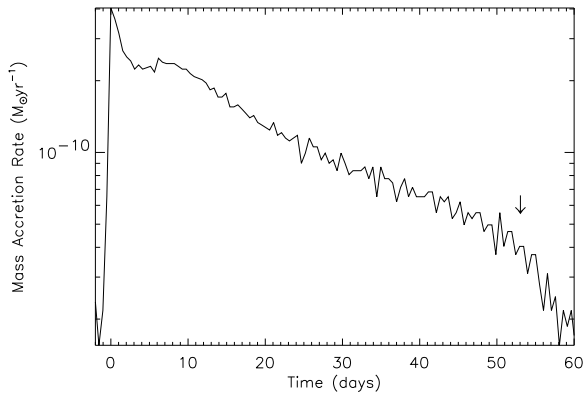,width=8cm}
\caption{Simulated outburst of A0620-003 with $\rm \eta$ reduced so that $\rm R_{h,max} = 0.35R_{disc}$. The
point at which $\rm R_h$ becomes smaller than $\rm 0.35R_{disc}$ is marked with an arrow.}
\label{ob2}
\end{figure}
\noindent

In the second simulation, shown in fig.~\ref{ob2}, the efficiency, $\rm \eta$, is reduced such that $\rm R_h$ 
is limited to a maximum value $\rm R_{h,max} = 0.35R_{disc}$. This allows us to mimic the effect of a 
self-shadowed outer disc. In this case, when the mass accretion rate starts to rise, only the inner part of the 
disc is irradiated. The accretion of gas in the irradiated region gives rise to the first peak in fig.~\ref{ob2}, 
which decays exponentially on a time-scale appropriate to the inner part of the disc. Meanwhile, the disc 
encounters the 3:1 tidal resonance and a two-armed spiral wave is launched in the disc. Gas in the outer, 
unirradiated regions  is accumulated in the spiral arms and the surface density rises. Eventually, 
$\rm \Sigma$ exceeds the critical limit for the viscosity to be increased. When this gas subsequently arrives 
at the primary, there is a rebrightening in the X-ray light curve. The second stage of the decay takes  place on
a longer time-scale, and the slope is correspondingly more gradual. The final, rapid drop to quiescence only 
takes place when $\rm R_h < 0.35R_{disc}$.

The two distinct regions, one irradiated and one unirradiated but subject to the disc instability, give rise to the
rebrightening. The light curve looks more like that of A0620-003, but differs in one important detail - the slope 
of the decay either side of the rebrightening is not the same. The observed similarity in the gradients of the 
decay either side of the rebrightening in A0620-003 suggests that the waves of mass which are producing these
parts of the outburst originate in the same region of the disc.
\begin{figure}
\psfig{file=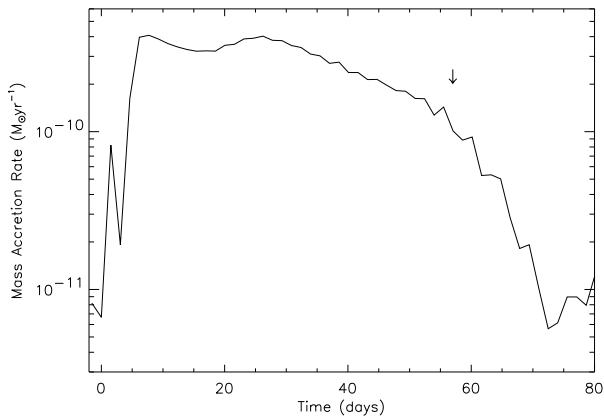,width=8cm}
\caption{{\bf Run A} : Simulated outburst of A0620-003 with $\rm \eta$ reduced so that 
$\rm R_{h,max} = 0.85R_{disc}$. The arrow marks the point at which  $\rm R_h$ becomes smaller than 
$\rm 0.85R_{disc}$.}
\label{ob1}
\end{figure}
\noindent

It is clear from the discussion in section 4.1 that $\rm R_h$ becomes much larger than the outer edge of the 
disc during an outburst of A0620-003. However, if a small part of the outer disc is shadowed from the radiation 
in some way, the rebrightening and the decay slopes could be reproduced even when $\rm R_h >> R_{disc}$.
In order to demonstrate this, a simulation was performed in which the efficiency limits $\rm R_h$ to a
value near, but not quite as large as $\rm R_{disc}$ (Run A). It is not our intention to suggest that the 
accretion  efficiency is extremely low, but to perform a simulation that mimics the shadowing of a small part 
of the outer disc during outburst. The resultant light curve is shown in fig.~\ref{ob1}, and this time the 
rebrightening and slopes are closer to the observations. Again, the final drop to quiescence only occurs when 
$\rm R_h < 0.85R_{disc}$.

\subsection{Varying the simulation parameters}

In order to assess the robustness of these results we performed run A using different SPH viscosities
and different particle numbers. A summary of the parameters in each run is given in table 1.
\begin{table}
\begin{tabular}{|c|c|c|c|}	\hline
Run & $\rm \zeta_{hot}$  & $\rm N_0$ & Figure\\	\hline
A & 1  & 49814 & 6\\
B & 1 &  24996 & 7a\\
C & 1 & 12588  & 7b\\
D & 1.5 & 49814 & 8a\\
E & 0.5 & 49814 & 8b\\	\hline
\end{tabular}
\caption{Summary of run parameters. $\rm N_0$ is the number of particles in the disc at the start of the 
simulation.}
\label{tab1}
\end{table}
\noindent
\begin{figure}
\psfig{file=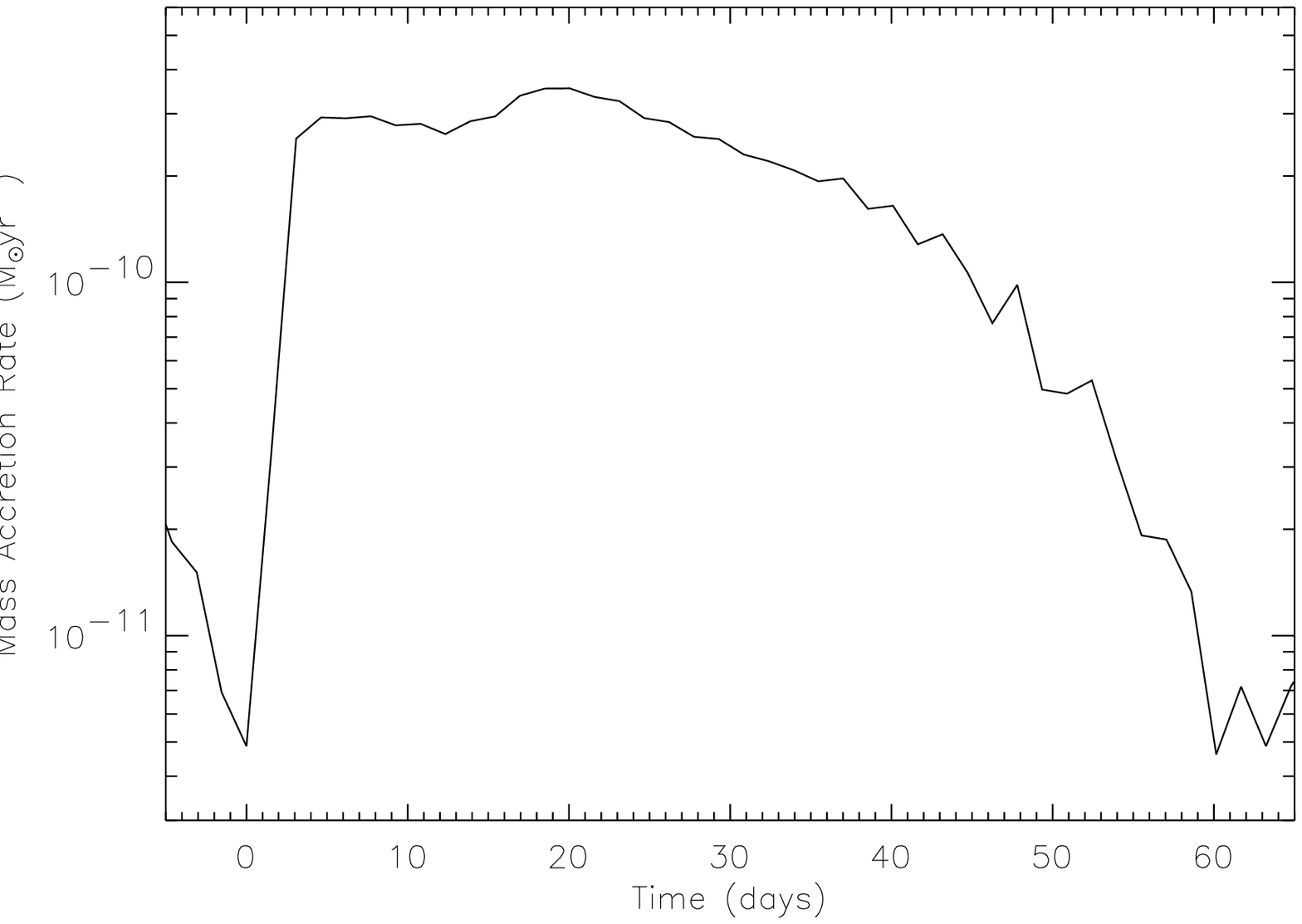,width=8cm}
\psfig{file=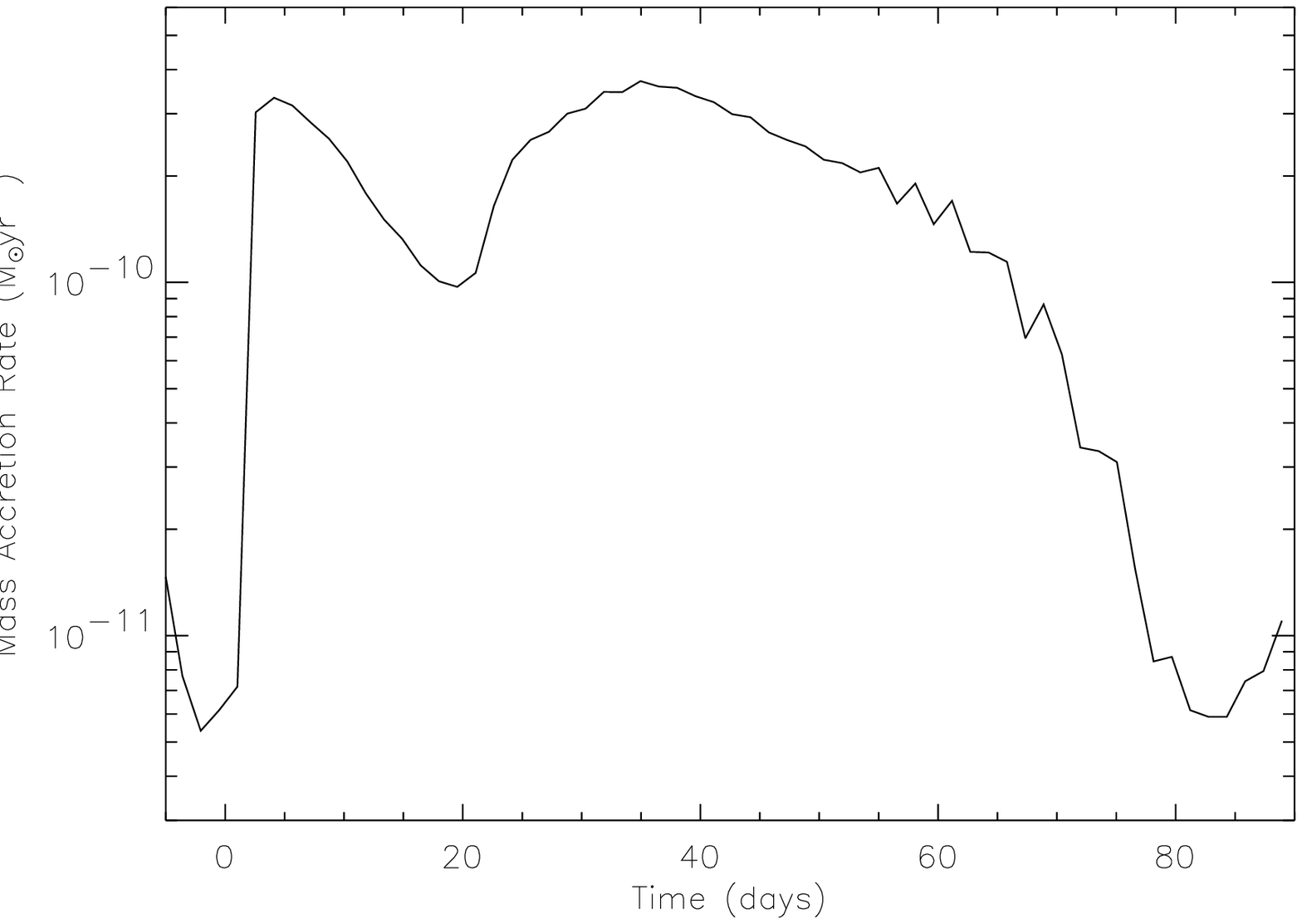,width=8cm}
\caption{Outburst simulations with fewer particles. Top : Run B. $\rm N_0 = 24996$ ; Bottom : Run C.
$\rm N_0 = 12588$.}
\label{fewpar}
\end{figure}
Firstly, the simulation was repeated with the same viscosity as before, $\rm \zeta_{cold} = 0.1, \zeta_{hot} = 
1.0$,  but with fewer SPH particles (runs B and C). Figure \ref{fewpar} shows the results with roughly one 
half and one quarter the number of particles of the orginal calculation. The greatest difference as particle 
number is decreased is the resolution of the primary maximum. With fewer particles, the spatial resolution 
in the inner part of the disc where the density is lower suffers to a greater extent than the outer part. 
Consequently, the accretion of particles from the inner disc leads to an under-resolved maximum in the 
accretion rate. Therefore, we expect the primary maximum to be even better resolved for a higher particle 
number, and the convergence between runs A and B is encouraging. The particle masses are scaled so that 
the total mass of the disc is the same in each simulation, so in run C the large drop in mass transfer rate 
before the rebrightening is caused by the resolution being very poor indeed - the accretion of just a few 
particles corresponds to a large response in accretion rate. The accuracy of this run should be judged 
accordingly.  
\begin{figure}
\psfig{file=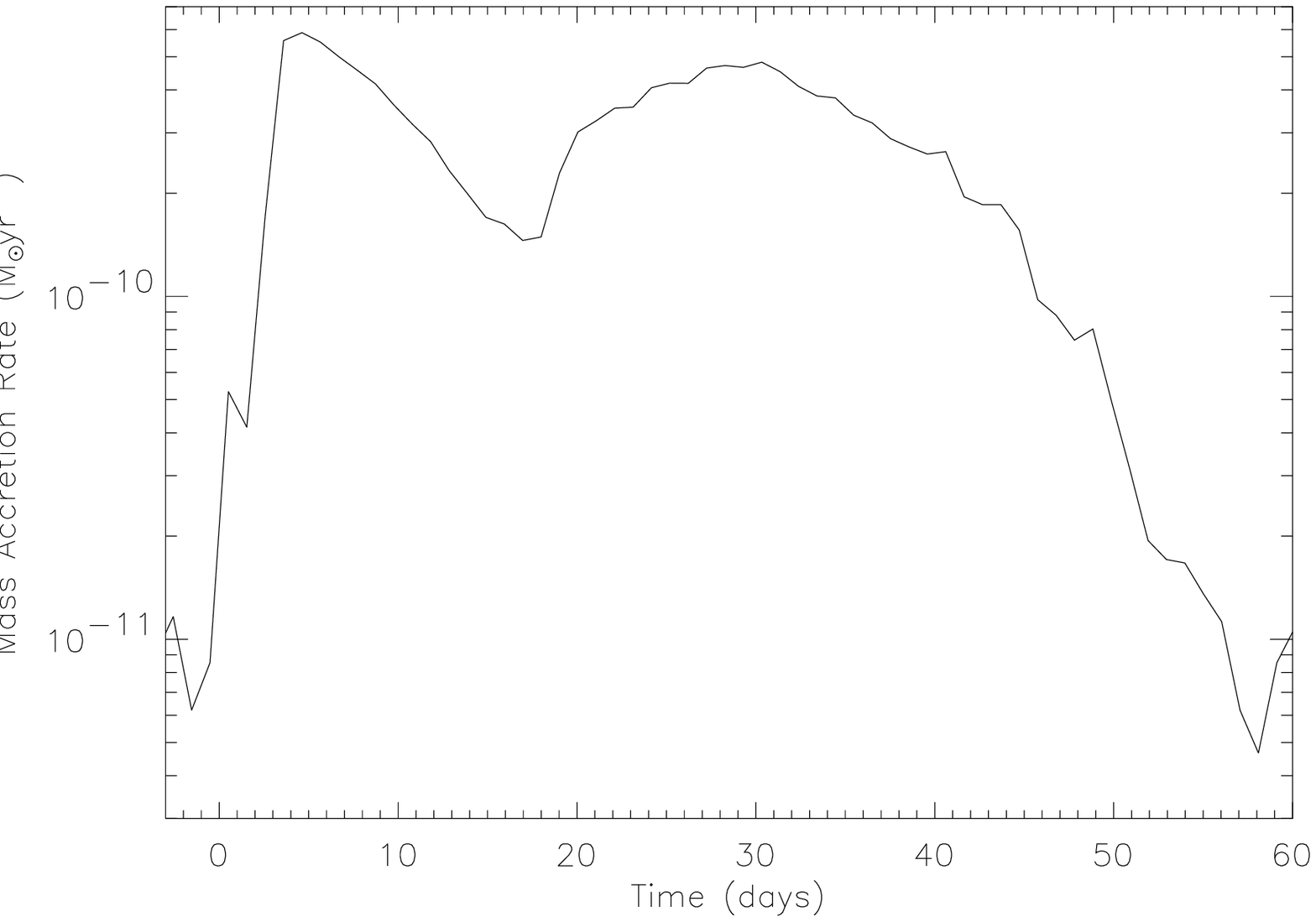,width=8cm}
\psfig{file=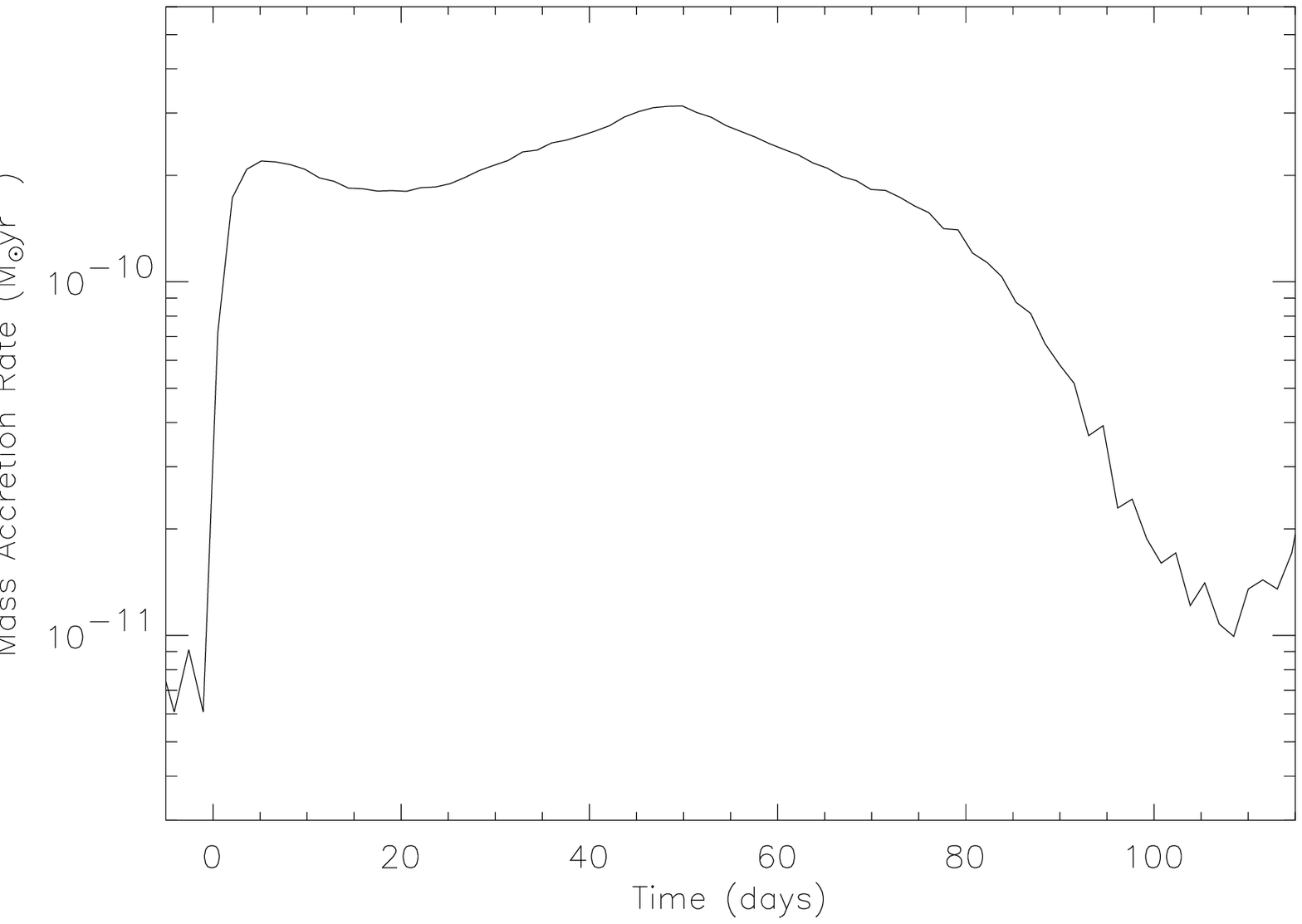,width=8cm}
\caption{Top: run D. $\rm \zeta_{hot} = 1.5$ ; Bottom: run E. $\rm \zeta_{hot} = 0.5$. In both runs, 
$\rm N_0 = 49814$. }
\label{zetanew}
\end{figure}
\begin{figure}
\psfig{file=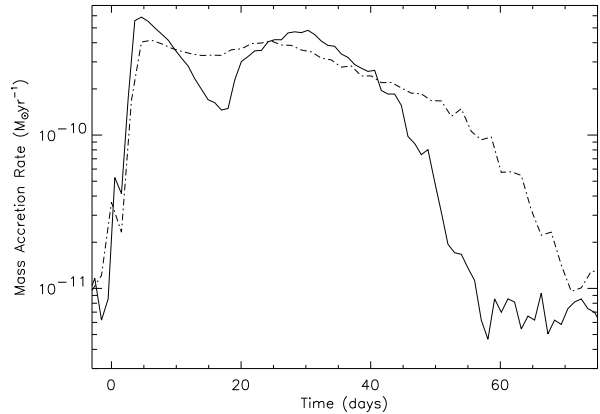,width=8cm}
\caption{Overlay of the outbursts in run A, with $\rm \zeta_{hot} = 1.0$ (solid line) and run D, with  
$\rm \zeta_{hot} = 1.5$ (dot-dashed line). The time delay between the onset of the outburst and the start of 
the rebrightening is the same in both simulations; however, the rise-time of the rebrightening is longer for the
lower viscosity.}
\label{compvisc}
\end{figure}

In runs D and E we keep the particle number the same as in run A, but vary the magnitude of $\zeta_{hot}$.
The resulting light curves are shown in fig. \ref{zetanew}. In both runs, the rebrightening starts at the same
time as it does in run A - between 15 and 20 days. However, the rise-time of the rebrightening is seen to
increse as $\rm \zeta_{hot}$ is decreased (fig. \ref{compvisc}). This suggests very strongly that the the 
rise-time of the rebrightening is governed by the viscosity, but the time delay from the start of the outburst to 
the rebrightening is independent of the viscosity. 

\section{Discussion}
In this section we discuss the applicability and limitations of these results to interpreting observations of 
A0620-003 and suggest a possible mechanism for the rebrightening. 
\subsection{Thermal structure}
We have already seen in section 4.1 that the SPH viscosity used in this isothermal scheme has a slightly 
different functional form to that of a standard $\rm \alpha$-disc.  Fig.~\ref{sphviscplot} shows that in
general the viscosity is slightly overestimated at small radii and slightly underestimated at larger radii.  It
should also be kept in mind that in the simulations, the irradiated region is indistinguishable from an
unirradiated region in the high-viscosity state. The question thus arises as to how good an approximation this 
is to a real irradiated disc, i.e. does the radiation alter the thermal structure? The magnitude of the 
optical depth, $\rm \tau$, and the origin of opacity in the disc is the key issue. $\rm T_{irr}$ can be quite 
sufficient to  dominate the effective temperature of the disc
\begin{equation}
\rm{T_{eff} = \left ( {3GM_1\mdot(R) \over 8\pi \sigma R^3} \right )^{1 \over 4}}~~~,
\label{teff}
\end{equation}
\noindent
but the vertical thermal structure of an irradiated disc will only be altered if $\rm T_{irr}$ also exceeds the
mid-plane temperature of the disc, given by $\rm {T_m}^4 \sim \tau{T_{eff}}^4$. King (1998) has
argued that $\rm \tau$  is of order 10 and consequently the vertical structure of the outer disc is nearly 
isothermal, with the sound speed dependent on  $\rm T_{irr}$ and not $\rm T_m$. The viscosity is therefore 
independent of the local value of  $\rm \Sigma$ and the diffusion equation
\begin{equation}
\rm {\partial \Sigma \over \partial t} = {3 \over R}{\partial \over
\partial R}\left\{R^{1/2}{\partial \over \partial R}\left[\nu \Sigma
R^{1/2} \right]\right\},
\label{diff}
\end{equation}
\noindent
leads to
\begin{equation}
\rm {\partial \Sigma \over \partial t} \sim \Sigma
\label{dsig1}
\end{equation}
\noindent
and the observed exponential decays. 

However, simulations by Dubus et al. (1999), in which the vertical structure is calculated self-consistently 
including the effects of irradiation by a point source in the orbital plane, suggest that $\rm \tau$ is of the order
$\rm 10^2 - 10^3$ and the structure is close to that of an $\rm \alpha$-disc. Indeed, for an 
$\rm \alpha$-disc with a Kramer's opacity law,
\begin{equation}
\rm \tau = 1200 \left( \frac{\alpha}{0.1}\right)^{-\frac{4}{5}} \dot M_{16}^{-\frac{1}{5}}
\label{opdepth}
\end{equation}
\noindent
where $\rm \dot M_{16}$ is the mass transfer rate in units of $\rm 10^{16} gs^{-1}$. In the more familiar 
case of an unirradiated accretion disc, the equation for vertical radiative energy transport is:
\begin{equation}
\rm{{4 \sigma T_m^4 \over 3 \kappa \Sigma} = {9 \over 8}\nu \Sigma {GM \over R^3}}
\label{bal}
\end{equation}
\noindent
and we can see straight away that although the opacity $\rm \kappa$ varies only slowly with 
$\rm \Sigma$ and $\rm T_m$ in much of the disc, the viscosity $\rm \nu$ is not independent of
$\rm \Sigma$. With the alpha prescription,
\begin{equation}
\rm{{\partial \Sigma \over \partial t} \sim \Sigma^{5 \over 3}}
\label{dsig2}
\end{equation}
\noindent
gives a power-law decay. 

Clearly, the magnitude of $\rm \tau$ is crucial to determining any change in the 
vertical thermal structure of the disc. It remains unclear how the energy is deposited in the vertical direction  
in an accretion disc, and whether the $\rm \alpha$-model is a valid description of the real processes involved.
If the $\rm \alpha$-model breaks down we have no way of calculating $\rm \tau$. However,  it remains
entirely possible that $\rm T_m > T_{irr} > T_{H}$ and the irradiation ionizes the disc, bringing about a 
change in viscosity, but does not alter the vertical thermal structure. The assumption that we make in the 
code is that the viscosity increases when $\rm T_{irr} > T_{H}$ and there is no change in $\rm T_m$ (i.e. the
sound speed remains constant).

We do not claim that our approach is a more  accurate representation than a 1D model that is 
thermodynamically self-consistent within the framework of the $\rm \alpha$-model. Rather, we use the 
advantages of  our scheme, particularly the two-dimensional nature of the code, to show that tidal forces 
(which are calculated self-consistently) can have a significant effect on the accretion flow in SXT discs and 
can account for the rebrightening observed in the X-ray light curves.

\subsection{Rebrightening time-scales}

The rebrightening in all the simulations begins 15-20 days after the onset of the outburst, while the peak of the
second maximum occurs at different times according to the viscosity. This suggests that the rise-time of each
rebrightening is controlled by viscous effects, but the time at which the rebrightening occurs is not. 
In each case, the accreted gas that is responsible for the rebrightening in the simulation originates from the 
outer (initially unirradiated) part of the disc. 
\begin{figure}
\psfig{file=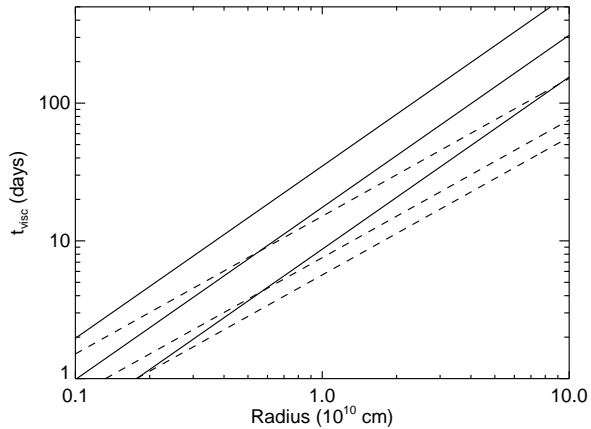,width=8cm}
\caption{Viscous time-scales. The solid lines are $\rm t_{visc}$ for an $\rm \alpha$-disc with 
$\rm \alpha = 0.1$ and $\rm M_1 = 4 \msol$. The upper solid line is for 
$\rm \dot M = 10^{-10} \msol yr^{-1}$, the middle $ \rm \dot M = 10^{-9} \msol yr^{-1}$ and the lower for  
$\rm \dot M = 10^{-8} \msol yr^{-1}$. The dashed 
lines show  $\rm t_{visc}$ in the SPH simulations, with $\rm c_s = 0.05 a\Omega_b$, $\rm M_1 = 4 \msol$, 
$\rm \zeta = 0.5$ (top), $\rm \zeta = 1.0$ (middle) and $\rm \zeta = 1.5$ (bottom).}
\label{tviscplot}
\end{figure}

The viscous time-scale for the outer part of a  Shakura-Sunyaev $\rm \alpha$-disc is longer than 50 days. 
The viscous time-scale is 
\begin{equation}
\rm t_v \sim {R^2 \over \alpha c_s H},
\label{tv1}
\end{equation}
\noindent
while the $\rm \alpha$-disc solution with a Kramer's opacity law gives
\begin{equation}
\rm t_v \sim 390~\left (\frac{\alpha}{0.1}\right )^{-{4 \over 5}}~\mdot_{16}^{-{3 \over 10}}~M_1^{1 \over 4}~R_{11}^{5 \over 4}~~~~d.
\label{tvisc}
\end{equation}
\noindent
where $\rm \mdot_{16}$ is the mass accretion rate in units of $\rm 10^{16} gs^{-1}$, $\rm M_1$ 
is the primary mass in solar masses and $\rm R_{11}$ is the radius in units of $\rm 10^{11} cm$.
This is plotted in fig. ~\ref{tviscplot} for the parameters of A0620-003, along with the SPH viscous time-scale.

Clearly, for the parameters of A0620-003 in outburst, the viscous time in the outer part of an 
$\rm \alpha$-disc is longer than 50 days. 
We have seen in section 3 that for such an extreme mass ratio ($\rm q = 0.07$), the 3:1 tidal resonance is 
easily accessible to the disc, and it is here that tidal effects start to affect the outburst. The tidal instability gives
rise to superhumps observed in the V-band light curve of CVs, and we have previously shown for a CV with 
$\rm q = 0.15$ that tidal effects prolong the outburst and drive eccentricity growth in the disc, with
the launch of a two-armed spiral wave (Truss et al., 2001). Exactly the same situation applies to the transient 
system studied here. Fig. 13 shows the evolution of the disc over run A - the spiral wave and the eccentricity 
are clearly evident during the outburst.  

The disc comes into contact with the 3:1 resonance soon after the onset of the outburst (fig. 13, top panel), but it takes 
some time for eccentricity to develop in the disc and the spiral waves to penetrate right down into the inner disc 
regions. Only then does the rebrightening occur (fig. 13, centre panel). By 
this time the disc has developed a significant  eccentricity and gas near the disc edge is spread over a range of 
different radii. So, what is the viscous time-scale near the disc edge? Consider the disc shown in the centre panel of 
fig. 13. Gas at the edge of the disc nearest the secondary is at a radius $\rm R \sim 0.3a$, while gas at the far edge is 
at $\rm R \sim 0.6a$. This factor of two in radius translates to a factor of two in viscous time-scale.  

In a real disc there is an additional factor which could drive accretion on a 
shorter-than-expected time-scale and produce the rebrightening.  Consider an idealised accretion disc with a 
Shakura-Sunyaev $\rm \alpha$ model structure. The equation of energy balance in the unirradiated part of 
the disc, \ref{bal}, is now:
\begin{equation}
\rm{{4 \sigma {T_m}^4 \over 3 \kappa \Sigma} = {9 \over 8}\nu \Sigma {GM \over R^3} + Q_{tide}.} 
\label{baltide}
\end{equation}
\noindent
where $\rm Q_{tide}$ is the energy dissipation due to tides. This can be written in the form given by Papaloizou
\& Pringle (1977):
\begin{equation}
\rm Q_{tide} =  (\Omega_{Kepler} - \Omega_{orb})T_{tide}
\label{qtide}
\end{equation}
\noindent
where $\rm T_{tide}$ is the tidal torque per unit area. In 1-D treatments of disc instabiliies, this is usually 
included as
\begin{equation}
\rm T_{tide} = CR\nu\Sigma\Omega_{orb}\left({R \over a}\right)^n
\label{torque}
\end{equation}
\noindent
\begin{figure}
\psfig{file=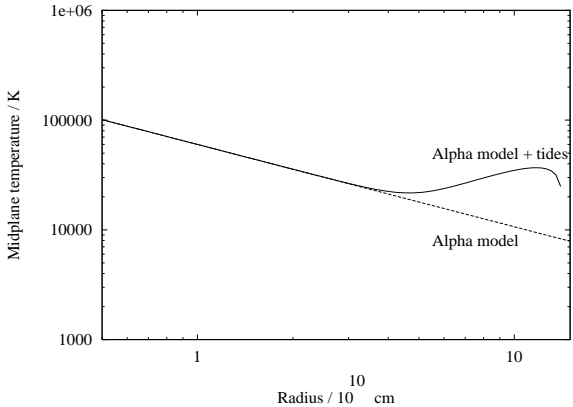,width=8cm}
\caption{The analytical solution for mid-plane temperature in an accretion disc with 
$\rm M_1 = 3.7 \msol$ and $\rm P_{orb} = 7.75 h$ as a schematic demonstration of the action 
of tidal heating. The dashed line is calculated from equation \ref{bal} and does not include the tidal term. The
solid line includes tidal torque calculated from equation \ref{torque} with n=5 and an arbitrarily large  value 
of C.}
\label{demo}
\end{figure}
\noindent
(Smak 1984; Hameury et al. 1998). There are two uncertainties in this equation: the value of C, which is a
numerically-determined constant, and the value of n, which determines how much of the disc is tidally
heated. Nevertheless, we can use this to demonstrate the effect of tides on the mid-plane temperature of
the disc. This is shown schematically for n=5 and an arbitrary value of C in fig. \ref{demo}. Clearly, in order to 
significantly impact on the mid-plane temperature, the tidal torque on the outer disc has to be both large and 
sustained.
\begin{figure}
\psfig{file=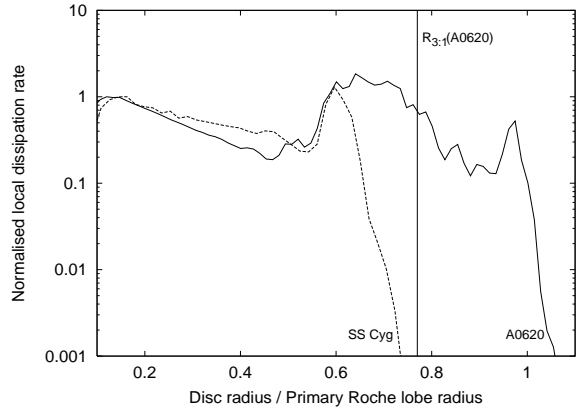,width=8cm}
\caption{Normalised energy dissipation rate averaged over one orbital period in outburst simulations of 
A0620-003 (solid line) and the cataclysmic variable SS Cygni (dashed line). SS Cygni is not subject to the 
tidal instability - the increased dissipation in a narrow region near the disc edge is due to the stream-impact 
hot-spot. In stark contrast, the tidal heating in A0620-003 is totally dominant, with a local dissipation rate over
an order of magnitude larger than the intrinsic disc dissipation rate acting over a very large area of the disc.
The dissipation rates have been normailsed to the rate at $\rm R = 0.15 R_L(1)$. The 3:1 radius of A0620-003
is marked with a vertical line.}
\label{comp}
\end{figure}

Figure \ref{comp} shows the energy dissipation rate in the disc averaged over one orbital period near the peak
of the simulated outburst in run A, along with a similar plot for a simulated outburst of the dwarf nova SS 
Cygni (with the same sound speed).  SS Cygni is not tidally unstable, since the disc never encounters the 3:1 
radius - the dissipation rate falls off with radius to the edge of the disc and rises in a well-defined narrow region
associated with the stream-impact hot-spot. However, in A0620-003, tidal forces produce the dominant 
contribution to  dissipation over a surprisingly large area of the disc. The tidal heating is spread over a range of 
radii because the disc is eccentric - over an orbital period the disc edge is not confined to a single radius. The 
dissipation rate in the outer regions of the disc is consequently about an order of magnitude higher than 
expected. We stress that this enhanced dissipation does not have any effect on the viscous
time-scale in the simulations : the viscosity is fixed at the value set in equation \ref{sphnu}.

\begin{figure}
\psfig{file=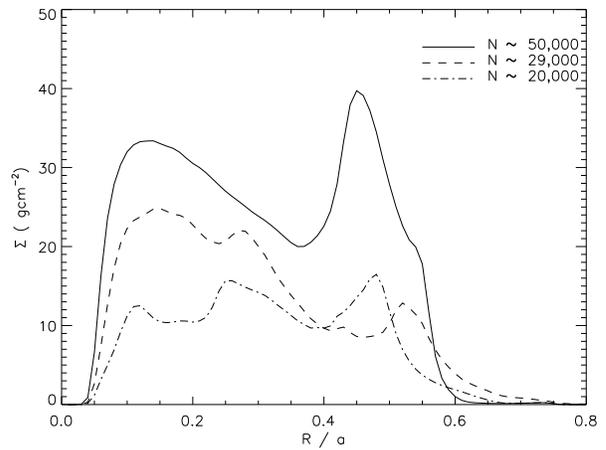,width=85mm}
\caption{Evolution of surface density throught the outburst in run A. N is the total particle number in the disc. 
The  curves correspond to t = 2 days (solid line), t = 30 days (dashed line) and t = 60 days (dot-dashed line). 
Most of the accreted mass goes into producing the first maximum in the light curve (fig. \ref{ob1}); the disc 
loses over 60\% of its mass in the first 20 days. The secondary maximum is caused by the arrival of mass 
initially near the disc edge (the large peak in the solid curve). }
\label{dens}
\end{figure}
\begin{figure}
\psfig{file=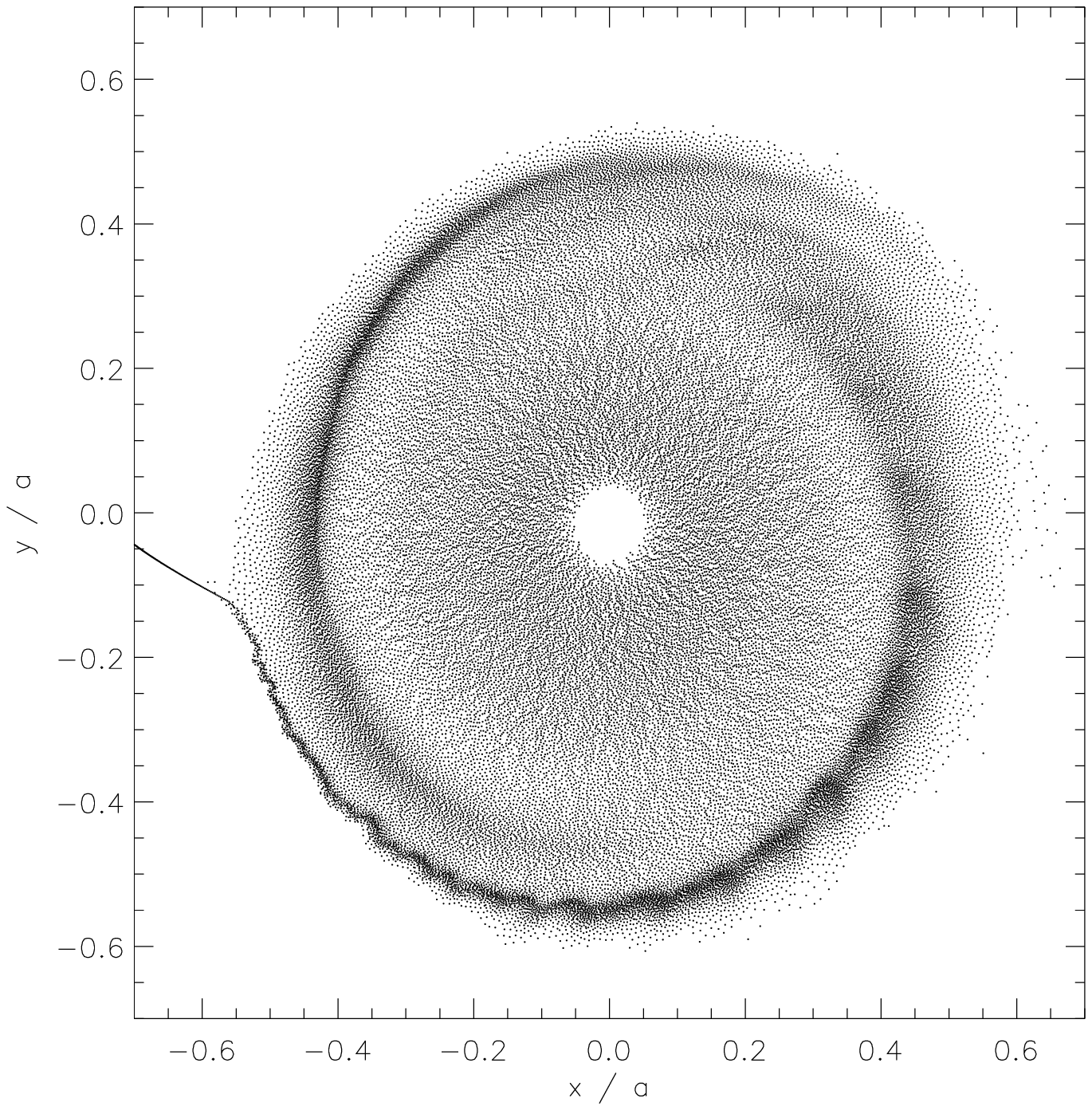,width=63.9mm}
\psfig{file=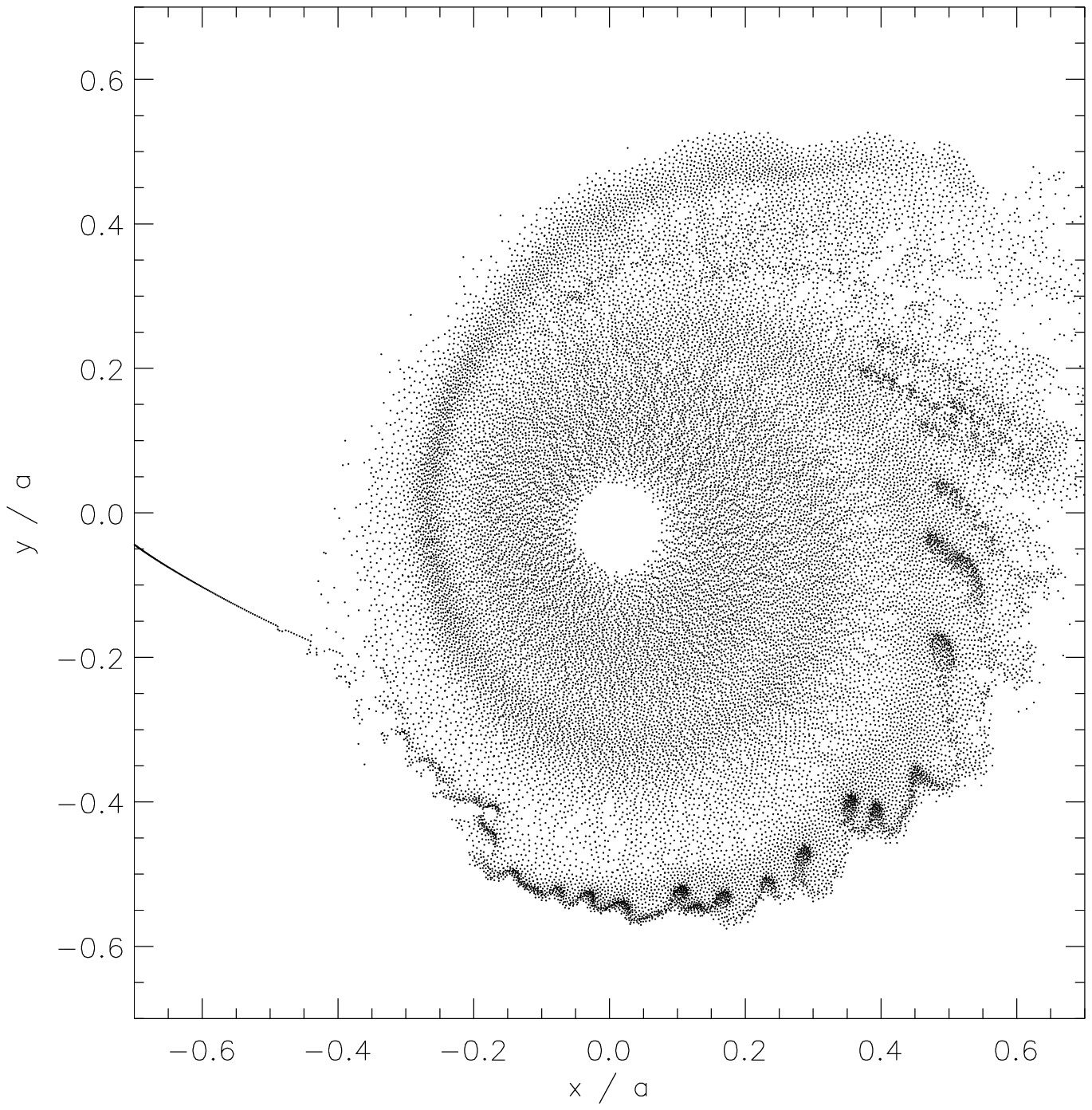,width=63.9mm}
\psfig{file=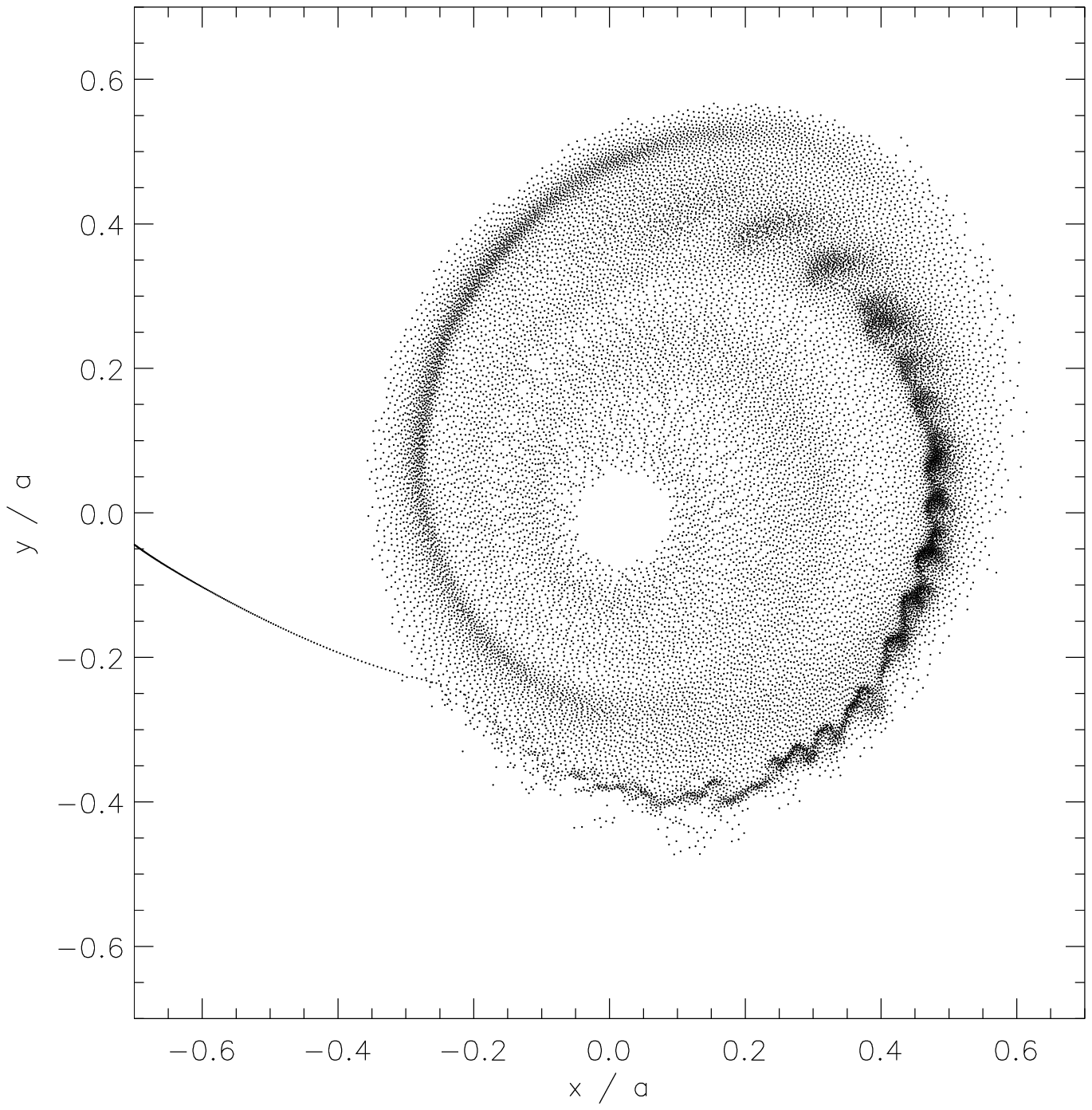,width=63.9mm}
\caption{Evolution of the disc through the outburst. The plots from top to bottom correspond to the same 
times as the three density profiles in figure \ref{dens} and are centred on the black hole primary. Particles 
are injected from the $\rm L_1$ point to the left of the plots; the inner part of each disc (white) is not modelled.
Near the peak of the primary maximum (top), $\rm R_h = 0.35a$ and the edge of the disc has encountered 
the 3:1 resonance (R = 0.47a). The rebrightening is associated with the arrival of gas in the tidally-induced 
spiral arm which penetrates into the inner regions of the disc (centre). Successive waves of material are 
accreted in this way as the outburst progresses, but the variations in the X-ray light curve diminish as the disc
is drained (bottom).}
\label{discs}
\end{figure}

Figures \ref{dens} and \ref{discs} show the evolution of surface density and particle position in the disc. The disc edge
very quickly encounters the 3:1 resonant radius and a two-armed spiral wave is launched (fig. \ref{discs}, top panel). 
At the peak of the outburst, material accumulates in the spiral arms in the unirradiated outer regions of disc (solid 
curve in fig. \ref{dens}),  while gas is the inner region is accreted. The surface density in the outer 'spike' of gas 
eventually surpasses the upper critical limit to trigger the disc instability.  As the outburst progresses, the disc
eccentricity grows.  The rebrightening is associated with the arrival of the initially unirradiated gas at the black hole.
This is not viscously spread on a time-scale that would apparently be appropriate to the outer disc - the spiral wave 
launched by the tidal instability penetrates the inner part of the disc and carries with it some of the gas in the outer 
disc (centre plot in fig. ~\ref{discs}).  Hence the rebrightening is caused by the accretion of a genuinely short pulse of 
gas. This penetration of the spiral arms into the inner regions of a hot disc has also been found in grid-based 
calculations performed by Stehle (1999).  Successive waves of material are accreted in this way as the outburst 
progresses. We find that $\rm \sim 60 \%$ of the disc mass is accreted in the primary maximum. This figure 
would be higher in a calculation in which a greater fraction of the disc is exposed to the radiation.

The enhanced dissipation in the outer regions of the disc is consistent with the findings of Blondin (2000), who used an
Eulerian scheme to show that tidal forces alone can produce a viscosity of the order of $\rm \alpha \sim 0.1$ in the outer
regions of an accretion disc. The effective viscosity decreased rapidly with radius, and only made a significant contribution 
in the outer half to two-thirds of the disc: exactly what we find with SPH in fig. \ref{comp}. It is still not absolutely 
clear that this will be sufficient to account for the 10 to 20 day time-scale for the rebrightening in A0620-003. Fig. 
\ref{tviscplot} shows that the viscous time-scale at the inner edge of the region of enhanced dissipation ($\rm R \sim 0.5 
R_L(1) \sim 6.5 \times 10^{10} cm$) is $\rm \sim 85$ days for a pure $\rm \alpha$-disc. However, Blondin (2000) and
Spruit (1987) note that the propagation of spiral waves  in a real disc is more likely than in an isothermal disc, because in 
a real disc the sound speed increases with decreasing radius ($\rm c_s \sim R^{-\frac{3}{8}}$). Hence, with a more 
realistic equation of state, the effective $\rm \alpha$ may be even higher in the inner disc. Furthermore, there is an
additional effect at work in the SPH simulations which was not found by Blondin - the development of a significant
eccentricity in the disc. These two effects may combine to make the  short time-scale of the rebrightening achievable.

\begin{figure}
\psfig{file=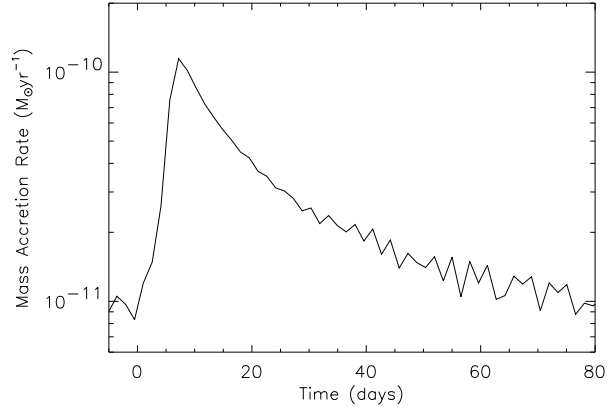,width=8cm}
\caption{Accretion rate as a function of time for a simulation of a transient with the parameters of A0620-003, except
that the mass ratio $\rm q = 0.6$. The disc does not become tidally unstable, and there is no rebrightening, just a long
tail. As before, $\rm \zeta_{hot} = 1.0$.}
\label{aonotide}
\end{figure}

In the SPH simulations, both sound speed and viscosity are fixed, so the rebrightening is caused by the arrival at the 
primary of a density wave associated with the spiral waves penetrating deep into the inner part of a highly eccentic disc. 
We perform one further simulation to show that the rebrightening is indeed a tidal effect. The same input parameters as
run A are used, with the exception that the mass ratio, $\rm q = 0.6$. At this mass ratio, the 3:1 resonance should not be 
accessible to the disc and therefore it cannot become tidally unstable. The result is shown in fig. \ref{aonotide}. As expected,
no significant eccentricity develops and there is no rebrightening: just a simple decay with a long tail.

In summary, then, there are two time-scales associated with the rebrightening: the growth time-scale of the tidal
instability in the disc, giving the time delay from the onset of the outburst to the start of the rebrightening, and the 
viscous time-scale of this tidally-stressed gas. In the simulations, the eccentricity of the disc causes gas initially near 
the disc edge to lie at much smaller radii. This explains why the rise-time of the rebrightening is so short : the gas that 
produces it is viscously spread on a time-scale much shorter than would be expected for gas originating near the edge 
of the disc. Undoubtedly, with a larger number of particles, the spatial resolution of the spiral shocks would improve 
and the simulated rebrightening would be even narrower.  We have also proposed that in a real disc (but not the
simulations) there is an additional factor that can increase the viscosity and make the rise-time of the rebrightening
much shorter than expected. This is the tidal heating of the disc, raising the mid-plane temperature.

\section{Conclusions}

We have simulated complete outbursts of soft X-ray transient systems, for the first time  in two dimensions. 
The  rebrightenings seen in the X-ray light curves of several systems can be explained by competition between 
the ionization instability, the effects of disc irradiation and tidal forces from the secondary star. Their 
appearance in the light curve is ultimately determined by the irradiated area of the disc, the location 
within the disc where the disc instability is first triggered and the growth time-scale of the tidal instability. The 
main findings of this work can be summarized as:
\begin{itemize}
\item{A rebrightening can occur if a small portion of the disc is shielded from the central X-rays 
in some way. This complements the conclusions of Dubus et al. (1999), who showed that a planar disc cannot 
be kept heated by irradiation from the central source, in contradiction to observations.}
\item{Tidal torques can be a dominant factor in the energy balance over a large area of the disc. Such an 
enahnced dissipation in a disc with the classical $\rm \alpha$-model structure will drive accretion on a 
shorter than expected time scale.}
\item{The time delay between the onset of the outburst and the start of the rebrightening is determined by the
growth time-scale of the eccentricity and spiral waves in the disc associated with the tidal instability.}
\item{The rise-time of the rebrightening is governed by the disc viscosity - but tidal eccentricity effects cause this
time-scale to be much shorter than expected for the arrival of gas from the disc edge.}
\item{No rebrightening is reproduced if the entire disc is irradiated.}
\end{itemize}

The main limitation of this approach is that the viscous evolution of the disc is governed by the local surface
density and not a local mid-plane temperature. The ultimate goal is to merge the current 1D thermodynamic 
state-of-the-art  with our 2D approach. The eventual extension to three dimensions is also particularly desirable 
in examining the beahviour of the tidal forces and the dynamics of the spiral arms.

There already exists a wealth of X-ray observations of transient sources, and we stress that although some 
aspects of the X-ray light curves are found to be common to many SXTs, they are all different to
some degree. In this work we have presented a mechanism for the outbursts which explains the origin of the 
rebrightenings observed in the X-ray light curves of some of these systems. 

Although one would not expect to observe superhumps in the X-ray light curve,
superhumps have been seen in optical light curves of SXT outbursts (O'Donoghue \& Charles 1996).
The superhump mechanism must be slightly different to that operating in cataclysmic variables, as the
optical emission from the disc is completely dominated by the reprocessed X-rays. Rather, the superhump
modulation can be attributed to the change in visible area of an eccentric disc over an orbital period (Haswell 
et al. 2001). This is exactly the scenario which we find in this work, and is a clear demonstration of the 
importance of using a two- (or preferably a three-) dimensional code able to take the tidal effects into 
account in a natural way.

\section*{Acknowledgments}
Research in theoretical astrophysics at the University of Leicester is supported by a PPARC rolling grant. 
Some of the simulations were performed on the UK Astrophysical Fluids Facility (UKAFF). Others were 
performed on the Theoretical Astrophysics Group's Linux Cluster which is supported by Advanced Micro 
Devices (AMD). The authors are grateful to the referee for extremely constructive comments and 
suggestions. MRT acknowledges a PPARC  studentship and the support of the William Edwards Educational
Charity.

\end{document}